\documentclass[10pt]{article}
\usepackage{authblk}
\usepackage{blindtext}

\usepackage[utf8]{inputenc} 
\usepackage[T1]{fontenc}    
\usepackage{hyperref}       
\usepackage{url}            
\usepackage{booktabs}       
\usepackage{amsfonts}       
\usepackage{nicefrac}       

\usepackage{multirow}
\usepackage{amstext}
\usepackage{amsmath}
\usepackage{amsthm}
\usepackage{amssymb}
\usepackage{bm}
\usepackage{wrapfig}
\usepackage{amsfonts, amsmath}
\usepackage[utf8]{inputenc}
\usepackage{graphicx}
\usepackage{bbm, comment}
\usepackage{color}
\usepackage{float}
\usepackage{wrapfig}
\usepackage{url}
\usepackage{MnSymbol,wasysym,marvosym,ifsym}
\usepackage[T1]{fontenc}
\usepackage[inline]{enumitem}

\usepackage{microtype}      
\usepackage{booktabs} 
\usepackage[ruled]{algorithm2e} 
\usepackage[tight]{subfigure}
\usepackage{tikz}
\usepackage{threeparttable}
\usepackage[utf8]{inputenc}
\usepackage{graphicx} 
\usepackage{float} 
\usepackage{pgfplots}
\usepackage{caption}
\usepackage{graphicx}
\usepackage{epstopdf}

\usepackage{tabularx}
\usepackage{pgfplotstable}

\usepackage{geometry}
\usepackage[numbers]{natbib}

\newtheorem*{theorem*}{Theorem}
\newtheorem{theorem}{Theorem}[section]

\newtheorem{definition}[theorem]{Definition}

\usepackage{xcolor}

\begin{document}

\title{SURPRISE! and When to Schedule It.}
\renewcommand{\thefootnote}{\footnote{footnote}}
\author[1,2]{Zhihuan Huang\thanks{Equal contribution}}
\newcommand\CoAuthorMark{\footnotemark[\arabic{footnote}]}
\author[1,2,4]{Shengwei Xu\protect\CoAuthorMark}
\author[3]{You Shan}
\author[1,2]{Yuxuan Lu}
\author[1,2]{Yuqing Kong\thanks{Corresponding author}$^,$\thanks{Supported by National Natural Science Foundation of China award number~62002001}}
\author[3]{Tracy Xiao Liu\thanks{Supported by the National Key Research and Development Program of China award number~2018YFB1004503}}
\author[4]{Grant Schoenebeck\thanks{Supported by (United States) National Science Foundation award number~2007256}}

\affil[1]{Department of Computer Science, Peking University}
\affil[2]{Center on Frontiers of Computing Studies, Peking University}
\affil[3]{School of Economics and Management, Tsinghua University}
\affil[4]{School of Information, University of Michigan}

\affil[1]{\texttt{\{zhihuan.huang, shengwei.xu, yx\_lu, yuqing.kong\}@pku.edu.cn}}
\affil[3]{\texttt{shany19@mails.tsinghua.edu.cn}}
\affil[3]{\texttt{liuxiao@sem.tsinghua.edu.cn}}
\affil[4]{\texttt{schoeneb@umich.edu}}
\renewcommand{\thefootnote}{\arabic{footnote}}
\maketitle



\begin{abstract}

Information flow measures, over the duration of a game, the audience’s belief of who will win, and thus can reflect the amount of surprise in a game. To quantify the relationship between information flow and audiences' perceived quality, we conduct a case study where subjects watch one of the world’s biggest esports events, LOL S10.  In addition to eliciting information flow, we also ask subjects to report their rating for each game. We find that the amount of surprise in the end of the game plays a dominant role in predicting the rating. This suggests the importance of incorporating when the surprise occurs, in addition to the amount of surprise, in perceived quality models. For content providers, it implies that everything else being equal, it is better for twists to be more likely to happen toward the end of a show rather than uniformly throughout.

\end{abstract}

\section{Introduction}

The live streaming industry has been burgeoning  around the world in recent years.  This includes live streaming games which, in turn, encompasses content like esports (e.g., League of Legends, Dota2, CS:GO, Apex Legends), sports games (e.g., football, tennis), and other games like chess, poker, and virtual casinos. Esports and its related brands occupy 24.2\% of the hours watched on Twitch.tv.\footnote{https://www.pwc.de/en/technology-media-and-telecommunication/digital-trend-outlook-esport-2020/media-broadcasts.html}
About 609 million people spent over 5 billion hours watching video game streams in 2016.\footnote{Based on Nate Nead's report \url{ https://investmentbank.com/esports-gaming-video-content/}}

Despite the popularity of these live shows, their quality varies significantly. We hypothesize an audience's perceived quality for such live streamed content is, in part, derived from the surprise in the content. One way to capture the effect of surprise is to solicit information flow delivered from the show. Before the game commences, the audience might have an imperfect idea of who will win. As the live game unfolds, the audience learns better about who the winner is likely to be.  In particular, the winner is clear by the time the game ends. Information flow measures, over the duration of a game the audience’s belief of who will win.  Intuitively, the surprise, measures how much information flow fluctuates over time. 

A key challenge is to quantify the relationship between the audiences' information flow and audiences' perceived quality. Prior studies either assume such relationship theoretically \cite{ely2015suspense} or use a statistical model to generate the theoretic information flow and indirectly measure audiences' perceived quality (e.g., by audience size) \cite{bizzozero2016importance,scarf2019outcome,buraimo2020unscripted}.  We instead elicit data directly from the audience to quantify the relationship and provide new insights for the development of such perceived quality models. Specifically, we elicit audiences' real-time beliefs to compute the amount of surprise in a game. We then study the relationship both between the amount of surprise and perceived quality and also the relationship between when the surprise occurs and perceived quality.

We design the Information Flow Elicitation Platform to collect the audiences' continuous beliefs and afterward ratings. Specifically, subjects watch live streaming games and update their beliefs for the games' outcomes as many times as they want. The platform monetarily rewards agents for their information flow reports in such a way that more accurate reports lead to higher payments. Subjects also rate the game quality afterwards. 

We use our platform to conduct a study targeting LOL S10.\footnote{The 2020 League of Legends World Championship is the tenth world championship for League of Legends, an esports tournament for the video game developed by Riot Games. It was held from 25 September to 31 October in Shanghai, China. There were 74 rounds of games in total and each game lasts for 30 to 40 minutes.}

\paragraph{Summary of our results.}
We find that the second half of the game has a larger amount of surprise compared to the first half and the amount of surprise at the end of the game has the strongest impact on the subjects' average ratings. Moreover, subjects' average ratings are significantly positively correlated with the games' surprise amount. Interestingly, the surprise amount in the first half of the game is negatively correlated with the average ratings, while this correlation in the second half is positive. One conjecture is that subjects overweight their watching experience in later time periods, which is not captured in prior studies. In other words, our results suggest that the perceived quality model should consider the time factor and the designers can use a better information revelation strategy such that the game is more likely to have a twist near the end. Additionally, we conduct robustness checks by considering alternative causes of perceived quality fluctuations, e.g., the favorite (home) team wins, and the results are consistent.

\section{Belief Curves, Median Curves, and Surprise}

In this section, we formally define the belief curve for each agent, and the aggregation of agent's beliefs into the information flow and median curve to compute the amount of surprise in a game.

We focus on the two-team competition setting. 

\paragraph{Belief curves and information flow.}

In game $g$, subject $s$ has a sequence of belief updates (the blue dots in Figure~\ref{fig:workflow}) chronologically $\{(t_0,p_0),(t_1,p_1),...,(t_n,p_n)\}$, where $n$ is the number of times that subject $s$ updates her belief in game $g$. Furthermore, $t_0=start_g$ shows that she reports her prior belief $p_0$ at the start of the game. Then she updates her belief from $p_0$ to $p_1$ at time $t_1$ and keeps updating her belief to the end. For convenience, let $t_{n+1}=end_g$. For all $0\leq i\leq n$, during period $[t_i,t_{i+1})$, subject $s$'s belief remains to be $p_i$. 

A subject $s$'s belief curve $p_g^s:[start_g,end_g)\mapsto [0,1]$ for a game $g$ represents her continuous belief throughout the game, where $p_g^s(t)$ is her belief for the winning probability of the blue team at time $t$ (Figure~\ref{fig:workflow}). The belief curve can be generated from her belief updates. Formally,

\begin{definition}[Belief curve]
Subject $s$'s belief curve is $p_g^s:[start_g,end_g)\mapsto [0,1]$ where \[
p_g^s(t):=p_i ~~ \text{ if } t\in [t_i,t_{i+1}) \text{ for all $0\leq i \leq n$ }
\]
\end{definition}

The \emph{information flow} is the collection of all the belief curves. 

\begin{figure}[!ht]
    \centering
    \includegraphics[width=1\linewidth]{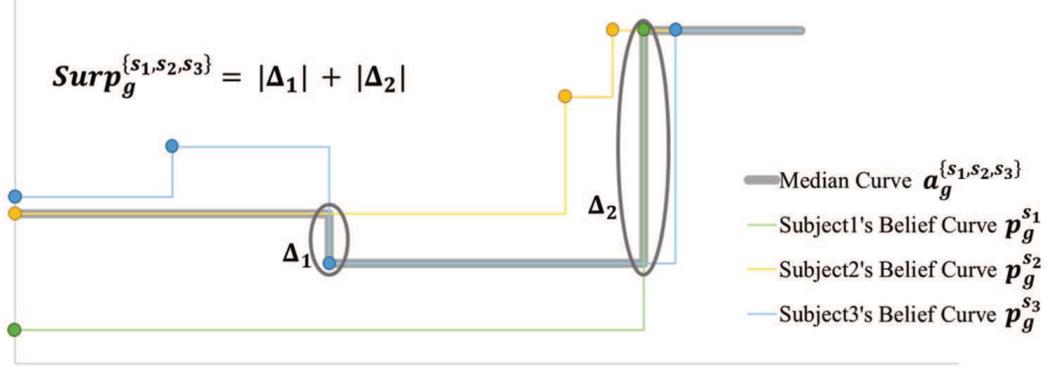}
    \caption{Surprise amount: we have three subjects $s_1,s_2,s_3$ whose belief curves are green, yellow and blue respectively. We aggregate their curves to a median curve which is the median of subjects' belief point wisely. The surprise amount is defined as the sum of changes, which is $|\Delta_1|+|\Delta_2|$.  }
    \label{fig:surpdef}
\end{figure}

\begin{figure}[!ht]
    \centering
    \includegraphics[width=0.99\linewidth]{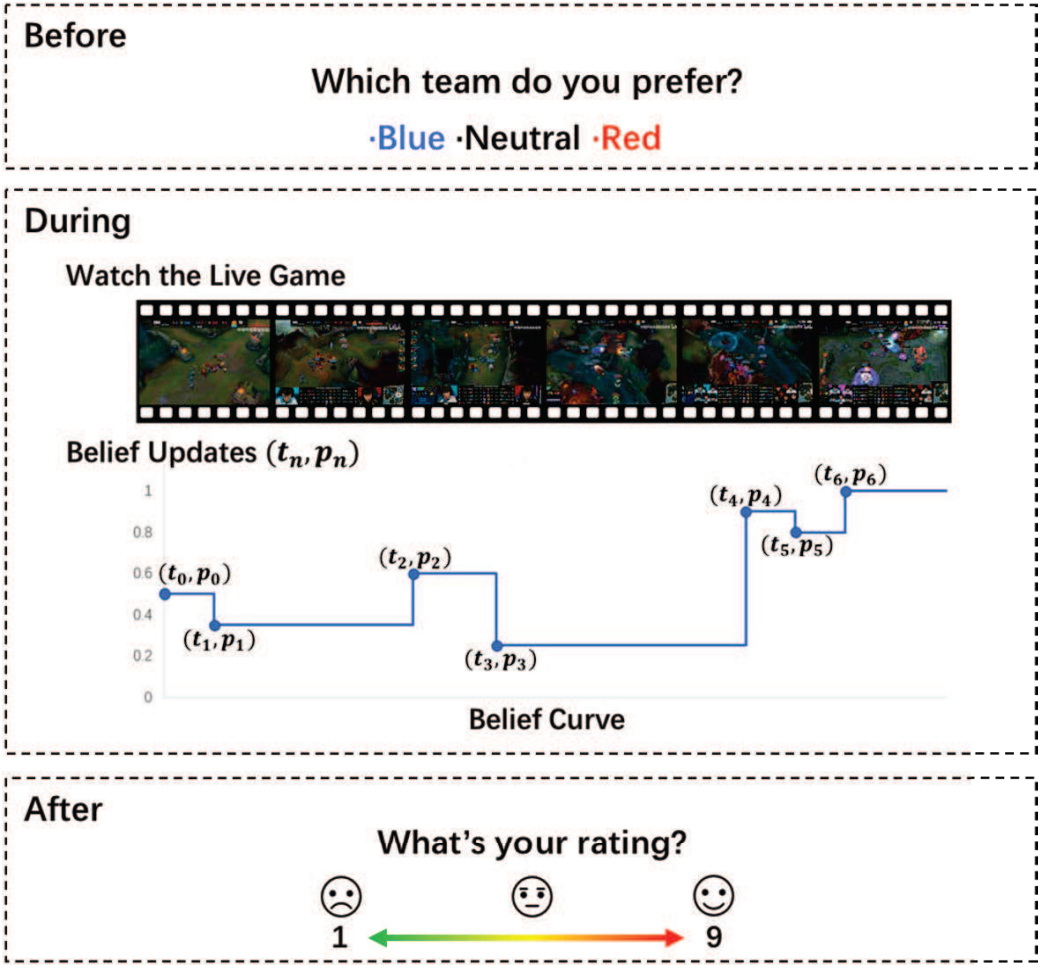}
    \caption{Workflow overview: we use a game in LOL S10 to illustrate the workflow. The game is between two teams, blue and red. We ask subjects their team preference before the game. Subjects view the game live and update their belief according to the game.\protect\footnotemark After the game, subjects rate the game.}
    \label{fig:workflow}
\end{figure}

\paragraph{Median curve.}  To reduce the bias caused by irrational agents who always report extreme beliefs (e.g., 0\%, or 100\%), we use the median curve to compute the surprise amount.  See Figure~\ref{fig:surpdef} for illustration of median curve and surprise amount.

\begin{definition}[Median curve]
For a game $g$ which is watched by a set $S$ of subjects, we define median curve $a_g^S: [start_g,end_g] \mapsto [0,1]$ as the median of the belief curve of all subjects in $S$ for game $g$, namely
\[
\forall t\in [start_g,end_g], a_g^S(t)=median(\{p_g^s(t)|s\in S\})
\]
\end{definition}

Figure~\ref{result:threeExamples} shows the median curves of three different games from our data set.  

\paragraph{Surprise.} Intuitively, if the median curve fluctuates severely, it suggests that this game has a high degree of surprise. Following \citet{ely2015suspense},  we define the amount of surprise as the sum of the change in the median curve.\footnote{This is seeming unrelated to the ``surprisal score'' sometimes used in Machine Learning.} Formally,

\begin{definition}[Surprise amount]

Given a curve which is a step function in $[x_0,x_{m+1}]$
\[
f(t)=\alpha_i ~~ \text{ if } t\in [x_i,x_{i+1}) \text{ for all $0\leq i \leq m$ }
\]
We define the surprise amount of this curve as 
\[ Surp(f):=\sum_{i=0}^m \left| \alpha_{i+1}-\alpha_{i}\right|\]
\end{definition}

$Surp_g^{S}:=Surp(a_g^S)$ is the amount of surprise in game $g$, which is the sum of absolute value of changes of the median curve $a_g^S$.\footnote{Since for all $s\in S$, $p_g^s(t)$ is a step function in $[start_g, end_g]$ of finite intervals, $a_g^{S}(t)$ is also a step function in $[start_g, end_g]$ of finite intervals.} We define $a_{g_1}^S$ as $a_g^S$ restricted to $[start_g,mid_g]$ and $a_{g_2}^S$ as $a_g^S$ restricted to $[mid_g,end_g]$ where $mid_g=\frac{start_g+end_g}{2}$. $Surp_{g_1}^{S}:=Surp\left(a_{g_1}^S\right)$ is the amount of surprise in the first half of game $g$ and $Surp_{g_2}^{S}:=Surp\left(a_{g_2}^S\right)$ is the amount of surprise in the second half of game $g$. 

\paragraph{Perceived quality vs. surprise.} We estimate $g$'s perceived quality by its average rating $r_g^S$ over all subjects $S$ who watch game $g$. To quantify the relationship both between the amount of surprise and perceived quality and also study the relationship between when the surprise occurs and perceived quality, we test the relationship between 1) game $g$'s surprise amount $Surp_g^S$ and its average rating $r_g^S$; 2) game $g$'s surprise amount in the first half $Surp_{g_1}^S$ and $r_{g}^S$; 3) game $g$'s surprise amount in the second half $Surp_{g_2}^S$ and $r_{g}^S$. 

\section{Data Collection Methods}

We first describe our Information Flow Elicitation Platform which was used to collect that data.  Second, we describe the data we collected.

\subsection{Information Flow Elicitation Platform}

A game is a competition between two teams, e.g., the red team vs. the blue team.   For each game, the study aims to collect three types of information from each subject: their team preference, their real-time belief of the blue team's winning probability, and their quality rating for the game. Specifically, there are three stages for each game: before, during, and after. Before the game, subjects report their preferences for the team. They also report their prior belief for the blue team's winning probability. During the game, subjects update their real-time belief of the winning probability whenever they want. After the game, they report their ratings for the game on a Likert scale, i.e., from 1 to 9, how much did you like the game?   

\begin{figure}\centering
  \subfigure[Before]{\includegraphics[height=.5\linewidth]{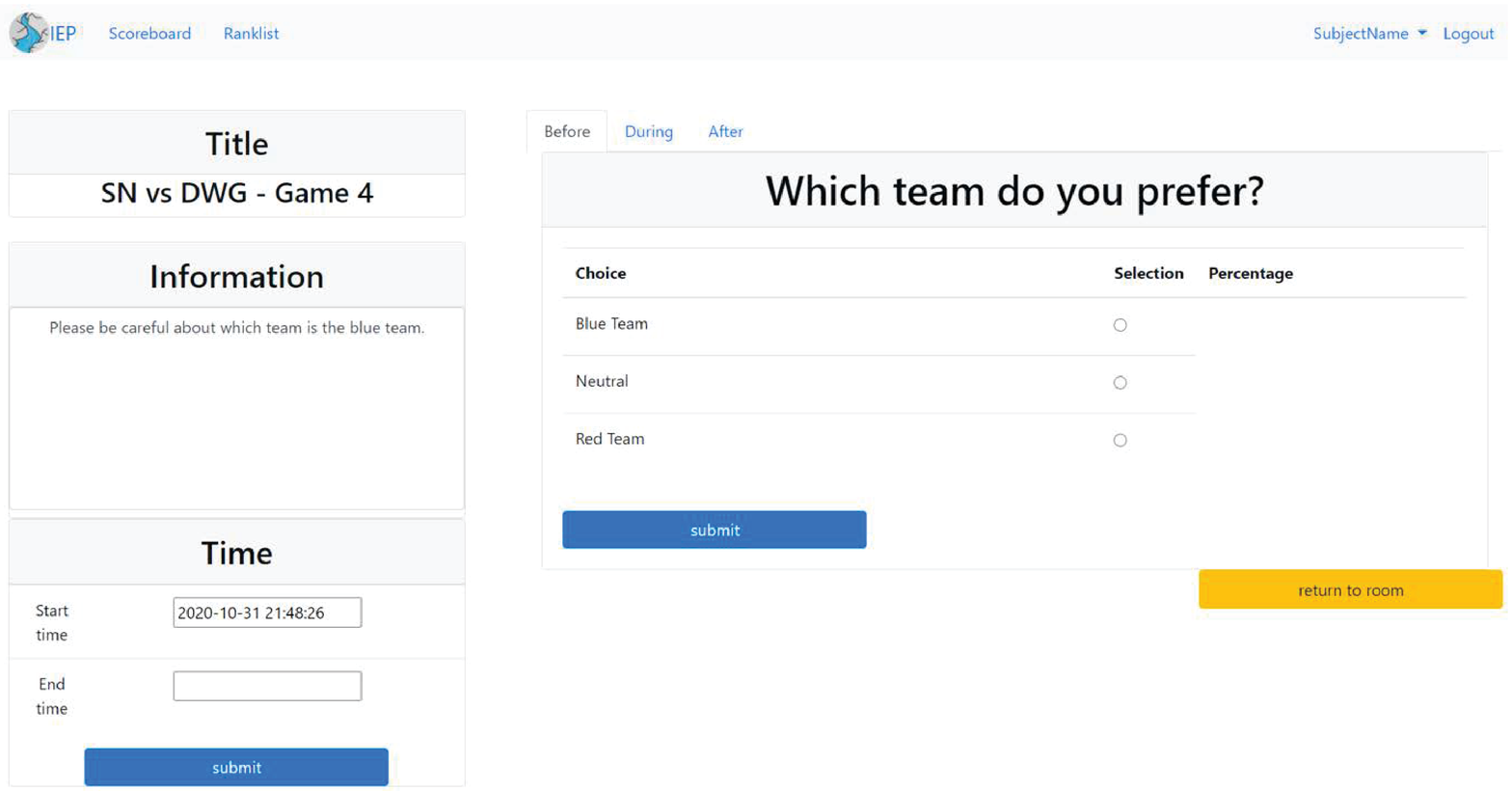}\label{platfrom:fig1:a}}
  \subfigure[During]{\includegraphics[height=.42\linewidth]{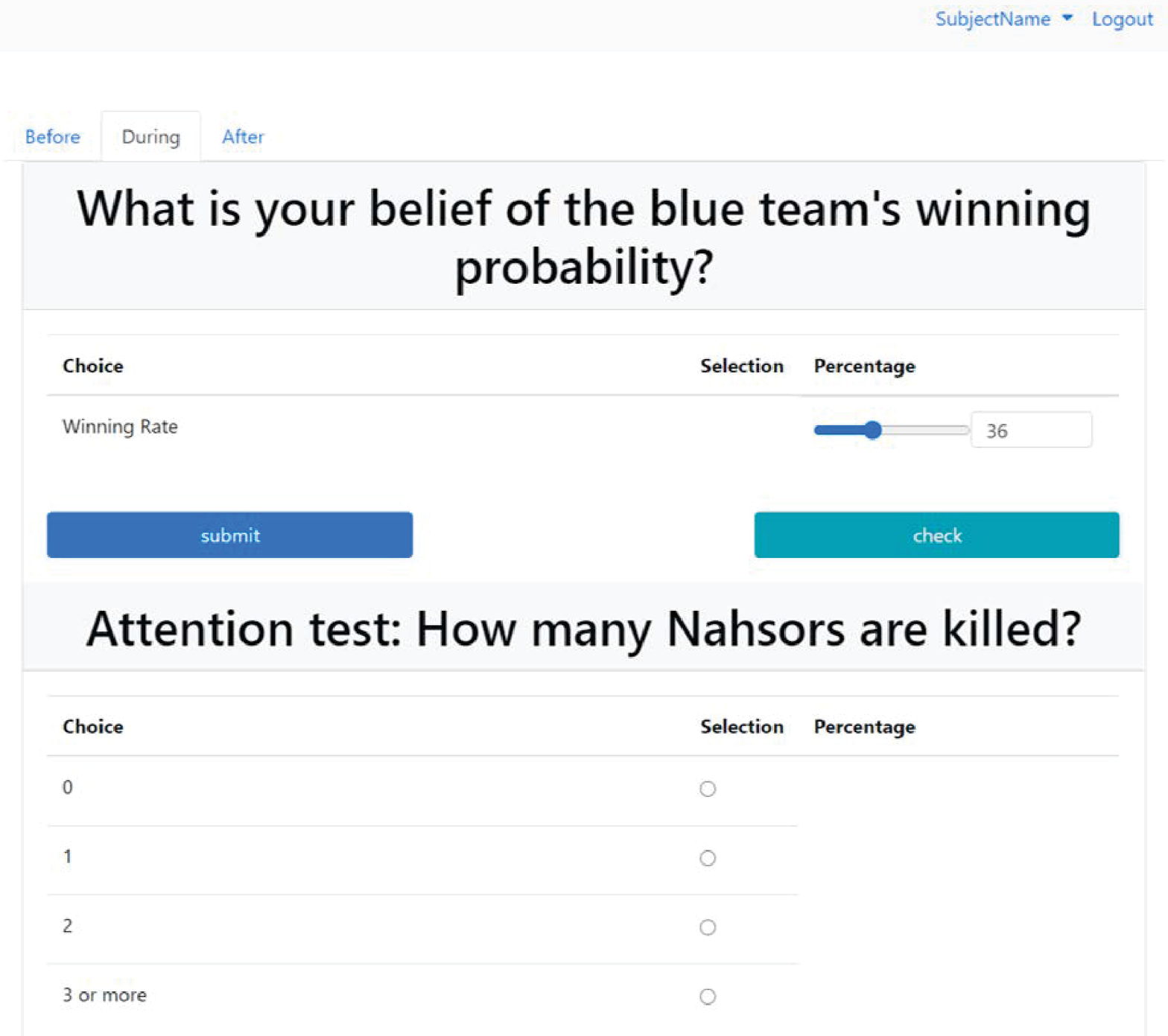}\label{platfrom:fig1:b}}
  \subfigure[After]{\includegraphics[height=.42\linewidth]{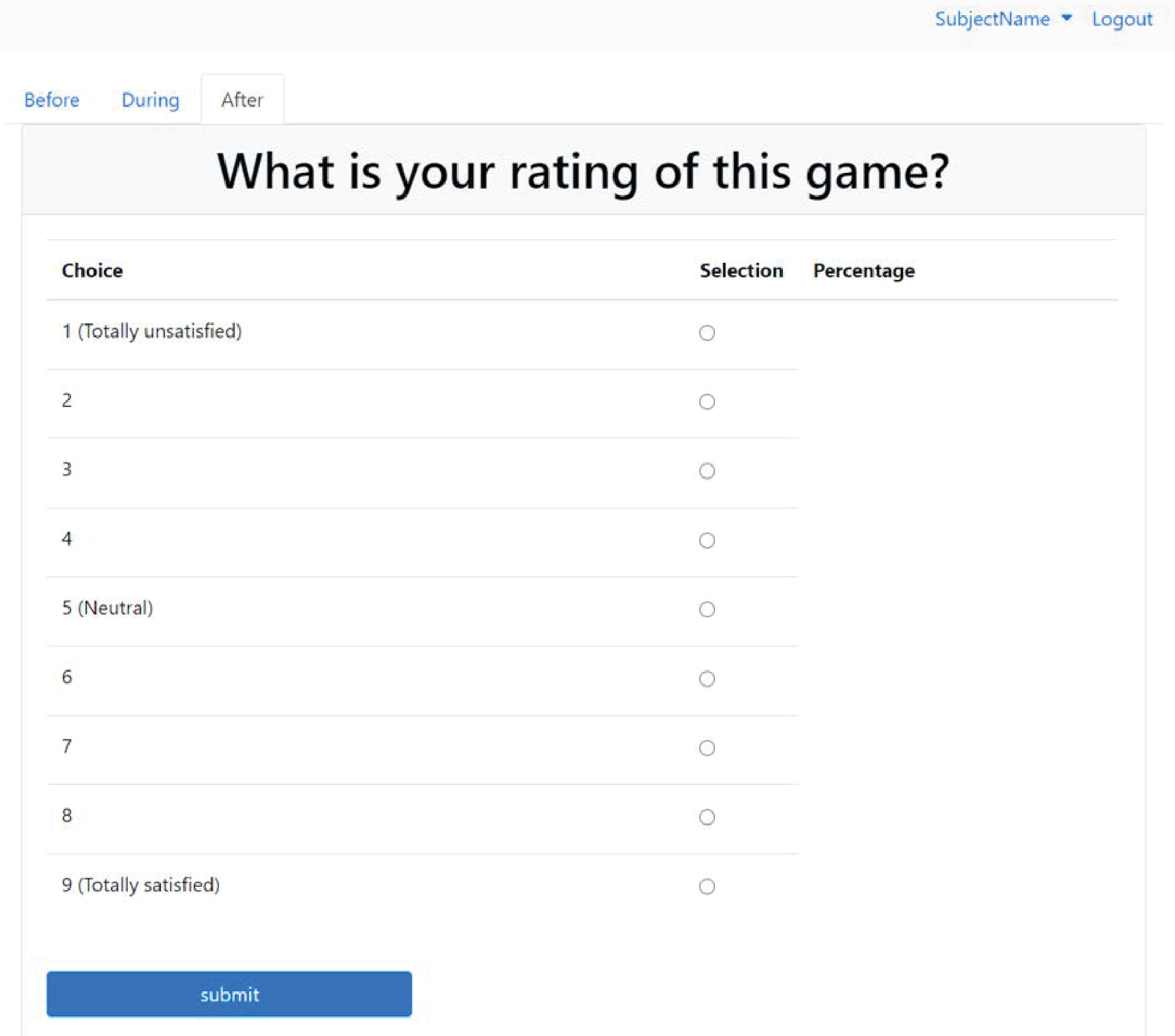}\label{platfrom:fig1:c}}
  \caption{Screenshots of our platform: the above figures are subjects' interface of our platform.} 
  \label{platfrom:fig1}
\end{figure}

\paragraph{Incentives.} For each game, subjects receive a monetary reward which depends on their overall prediction accuracy. To measure the overall prediction accuracy, we use the quadratic scoring rule \cite{Brier1950VERIFICATIONOF,properscoringrule2007} to measure the prediction accuracy at every $t$ and integrate the quadratic score over $[start_g,end_g]$. 

\footnotetext{The screenshots of the game is from LOL S10's live streaming platform, https://www.bilibili.com/ }

Formally, each subject receives a score which depends on her belief curve. When the game ends, the outcome $o_g$ for the blue team  is either $0$ or $1$. Subject $s$'s quadratic score at time $t$ is $1-(p_g^s(t)-o_g)^2$.  The overall quadratic score of subject $s$ is:
\[
Score(p_g^s)=\frac1{end_g-start_g}\int_{start_g}^{end_g}(1-(p_g^s(t)-o_g)^2) \text{dt}
\]

For example, we consider a game where the starting time is 00:00, the ending time is 00:50, and the red team wins in the end. A subject reports her prior belief 40\% for the winning probability of the blue team at the beginning. Then she updates her belief to 80\% at 00:25, 50\% at 00:30, and 0\% at 00:40. Her score will be $[(1-0.4^2)\times(25-0)-(1-0.8^2)\times(30-25)-(1-0.5^2)\times(40-30)-(1-0^2)\times(50-40)]\times (1/50)=0.86$.

For subject $s$, at every time $t$, the expected quadratic score is maximized when $p_g^s(t)$ is her true belief at time $t$. The expected score is maximized when $\forall$ $t$, $p_g^s(t)$ is her true belief at time $t$. Thus, our score is incentive-compatible. However, this leads to non-fixed cost. To fix the budget, following \citet{LAMBERT2015389}, we calculate the average score over all subjects in game $g$, $\overline{Score}_g$. Subject $s$'s reward is then \[(1+Score(p_g^s)-\overline{Score}_g)\frac{B}{M_g}\] where $M_g$ denotes the number of subjects in game $g$. With the aforementioned reward, the total reward for the game is fixed to $B$. Moreover, the reward is always non-negative and has the same incentive properties as the original score.

\subsection{Datasets}\label{sec:ourdata}

\paragraph{League of Legends.}

\begin{figure}[!ht]\centering
  \subfigure[Likelihood that G2 beats SN]{\includegraphics[height=.30\linewidth]{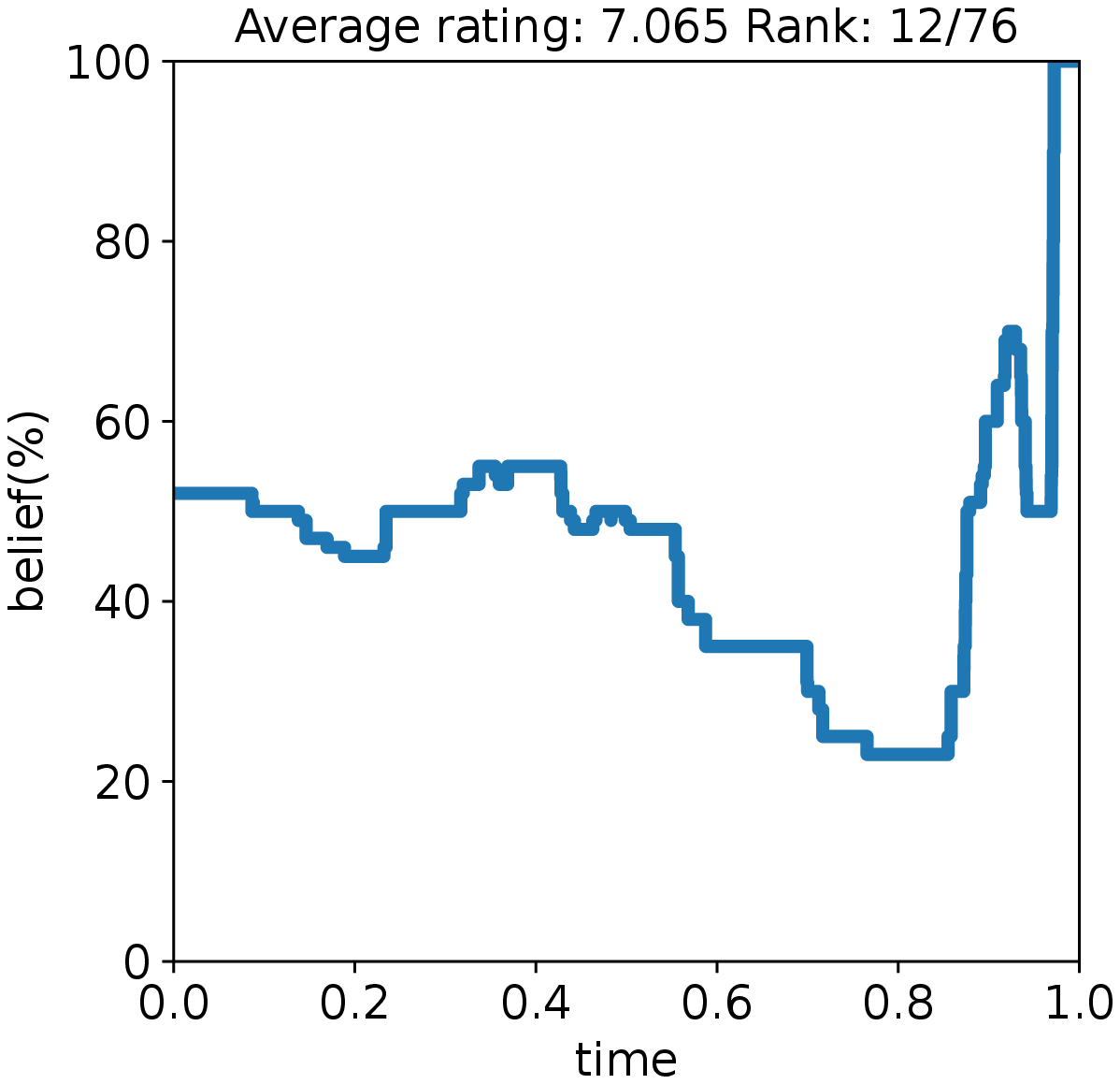}\label{result:threeExamples:a}}
  \subfigure[Likelihood that DWG beats PSG]{\includegraphics[height=.30\linewidth]{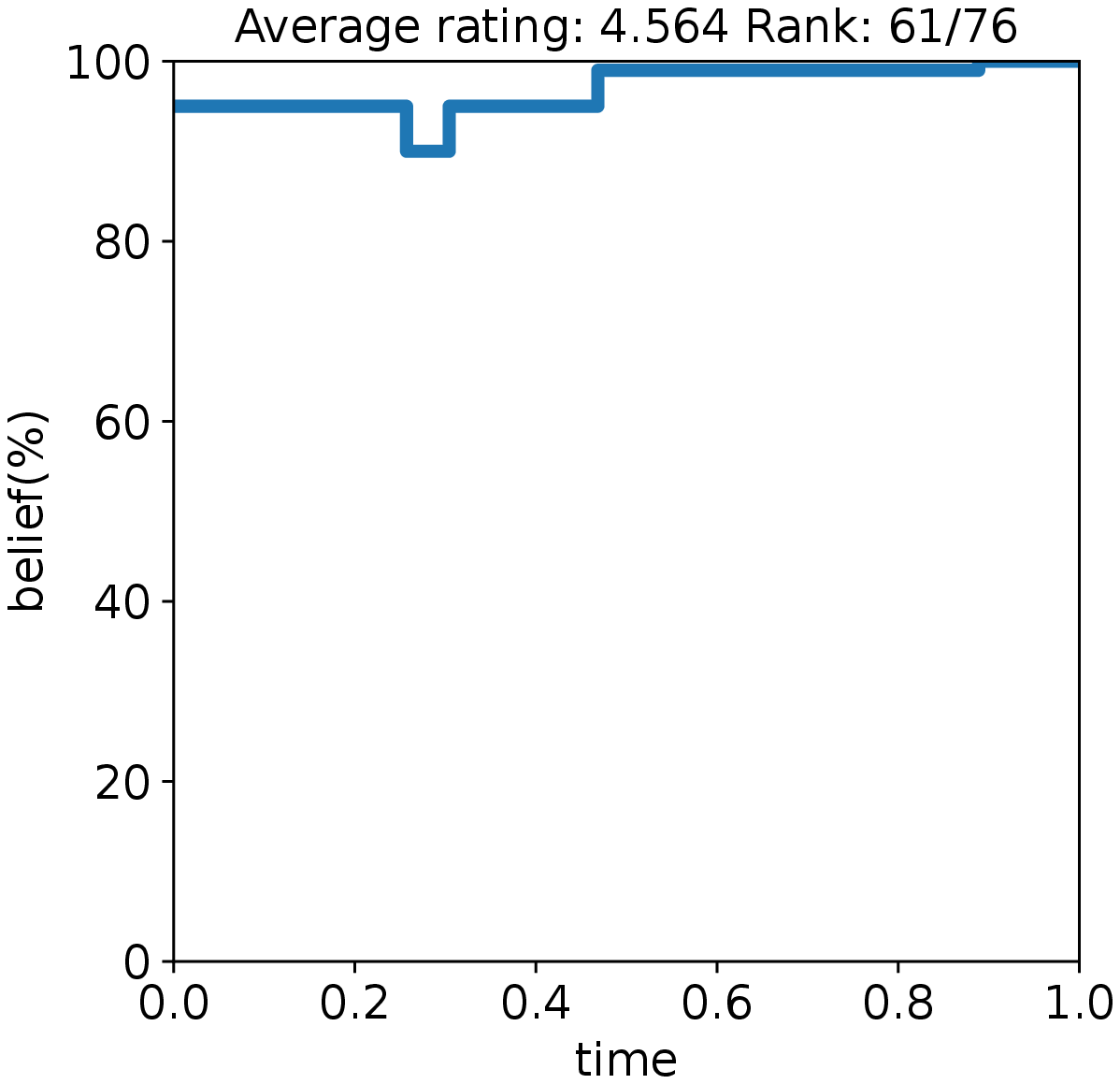}\label{result:threeExamples:b}}
  \subfigure[Likelihood that UOL beats DRX]{\includegraphics[height=.30\linewidth]{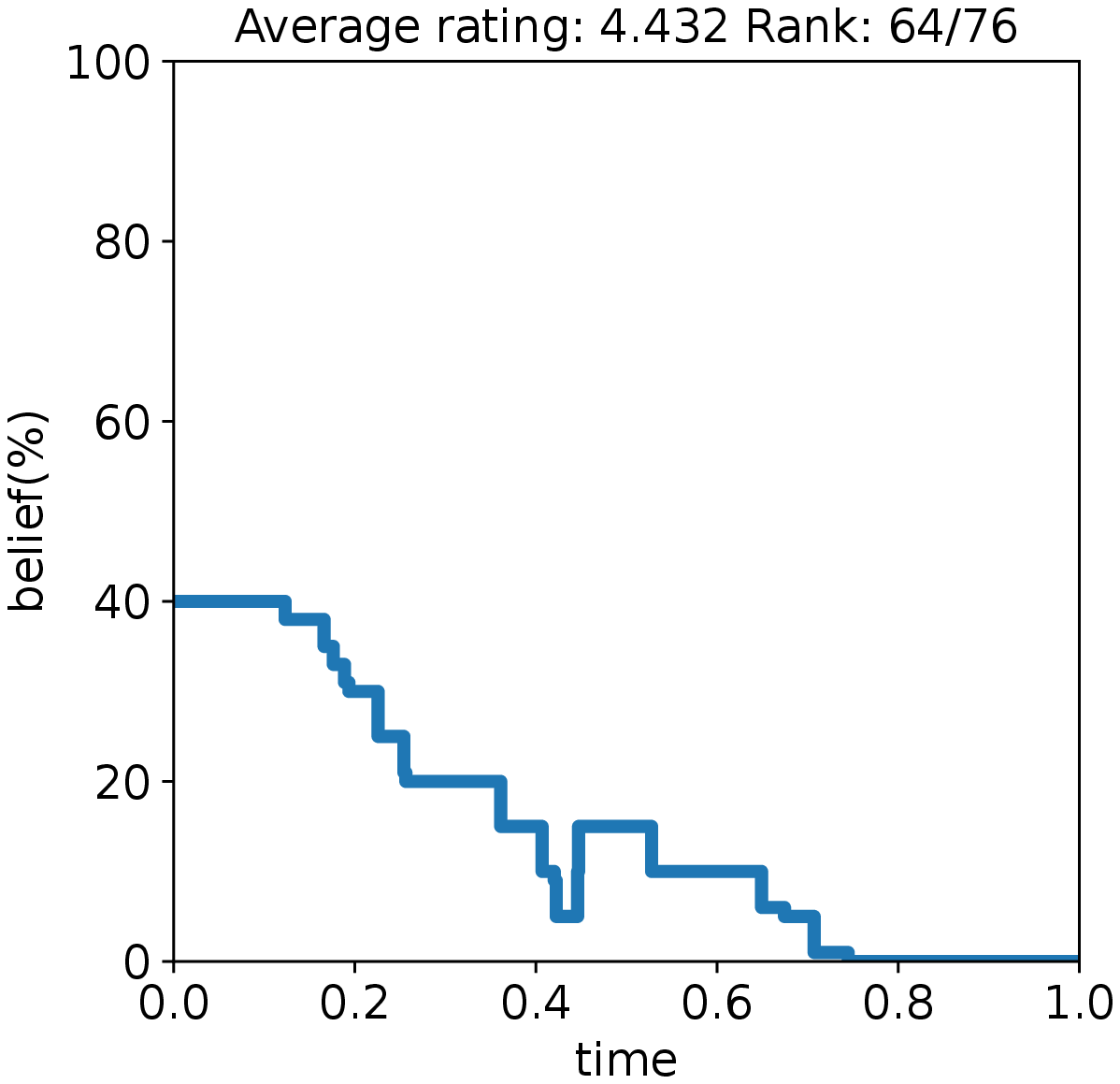}\label{result:threeExamples:c}}
  \caption{Median curves of three games in LOL S10: the figures above shows the median curves of three games with different ratings. The game in (a) has a very high rating: rank 12 among all 76 games. This game is between two well matched teams. There are several reversals in the game. The game in (b) has a low rating (rank 61/76); DWG is the champion team, and PSG is a weak team.  The subjects are confident that DWG will win in the beginning, and the outcome fulfills their expectation. The game in (c) also has a low rating (rank 64/76), UOL is slightly weaker than DRX. By the middle of the game, DRX has taken control and the second half has no big surprises.}
  
  \label{result:threeExamples}
\end{figure}

League of Legends is a free 5v5 online MOBA (multiplayer online battle arena) game created and published by Riot Games. 
The goal of the teams is to destroy the enemy team's base. The match ends immediately after one teams' base is destroyed.

\paragraph{Data Properties.}

We use our platform to conduct a study for LOL S10 which consisted of 76 individual games. We recruited 107 subjects from top Chinese universities.
For each game, a link to participate was sent out to all the participants.  Subjects could participate in as many or as few games as they like.  Additionally, we did not restrict the number of agents that signed up for each game.

We obtained 4,566 observations in total, where an observation consisted of one particular subject participating in one particular game.  5 subjects participated in all 76 games. 3 subjects of them only participate once.  The average number of games that a subject participated in was 42.67. 

\paragraph{Exploratory Data Analysis.}

The average score for our subjects in each game was 0.817. The average payment for our subjects in each game was 10.26 CNY (about \$1.58 USD), yielding a total payment of 46,850 CNY (about \$7,230 USD). 

Moreover, the median frequency for belief updating is 5 and the average is 5.87. 68\% subjects are majoring in STEM. All subjects report that they have experience watching LOL Live.

For each game, we can measure the number of subjects, the average rating, the duration, the peak time, the surprise in the first half and the second half, the peak surprise, the end surprise and the overall surprise.  The \emph{peak time} measures the most surprising time in the game which we define as the middle of the time interval of 2.5 minutes that has the maximum amount of surprise. The \emph{peak surprise} is defined as the surprise amount generated in the peak time. The \emph{end surprise} is defined as the surprise amount generated in the last 2.5 minutes.

\begin{table}[!ht]
	\centering
	\begin{tabular}{|p{3cm}<{\centering}|p{1.2cm}<{\centering}|p{1.2cm}<{\centering}|p{1.2cm}<{\centering}|}  
		\hline
		& average & min & max\\
		\hline
		number of subjects & 59.974 & 28 & 83\\
		\hline
		average rating   & 5.709 & 3.600 & 8.235\\
		\hline
		duration (min)  & 32.039 & 18.817 & 45.317\\
		\hline
		peak time (min)  & 23.950 & 2.600 & 44.042\\
		\hline
		1st half surprise  & 0.262 & 0.040 & 0.675\\
		\hline
		2nd half surprise  & 0.531 & 0.010 & 1.445\\
		\hline
		peak surprise  & 0.278 & 0.090 & 0.790\\
		\hline
		end surprise  & 0.162 & 0 & 0.725\\
		\hline
		overall surprise  & 0.793 & 0.150 & 1.750 \\
		\hline

	\end{tabular}
	\caption{Summary statistics of our data}
	\label{table:basic}
\end{table}

Table \ref{table:basic} displays the average, minimum, and maximum of each of these quantities. Note that on average, the surprise in the second half is twice the surprise in the first half. 

\begin{figure}[!ht]\centering
  \subfigure[The number of subjects in games]{\includegraphics[width=.48\linewidth]{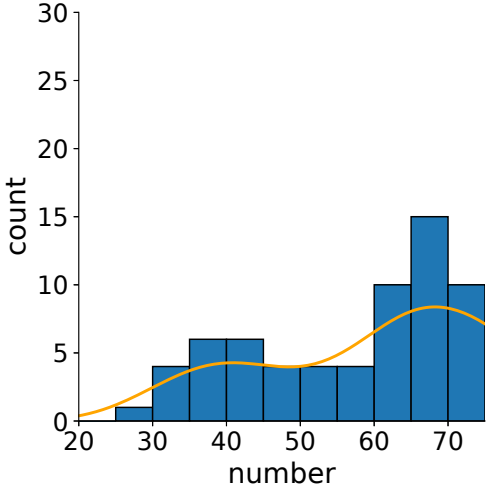}\label{result:subjects}}
  \subfigure[Average game rating]{\includegraphics[width=.48\linewidth]{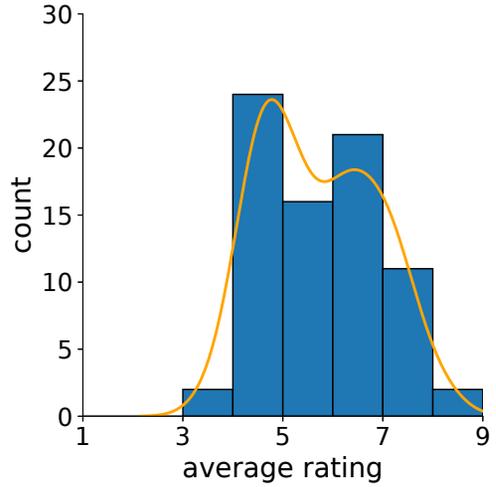}\label{result:average rating}}
  \subfigure[Length of the game]{\includegraphics[width=.48\linewidth]{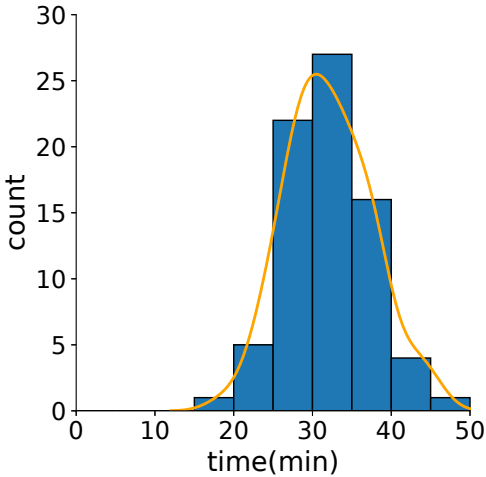}\label{result:length}}
  \subfigure[Time that peak occurs]{\includegraphics[width=.48\linewidth]{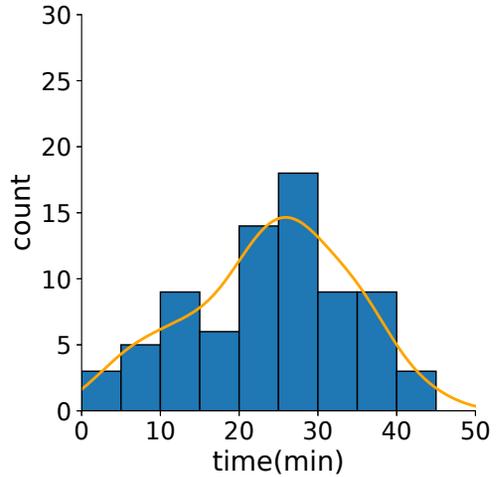}\label{result:peak time}}
  \caption{Histogram of multiple statistics over all games: a) number of participating subjects; b) the average ratings; c) duration  d) the peak times (when surprise is the highest). We also draw the kernel density estimation curve of these histograms.}
  \label{result:data}
\end{figure}

Figure~\ref{result:data} shows a histogram of the first four of these  quantities. Observe that the most frequent peak times are between 20 to 30 minutes.  This corresponds to a key part of the matches, killing the first dragon (Baron Nashor), which appears exactly at the 20th minute of the match, and is tyfiguresally killed between the 20 and 30 minute mark and grants the successful team a lasting advantage.

\begin{figure}[!ht]\centering
  \subfigure[First half vs. second half]{\includegraphics[width=.48\linewidth]{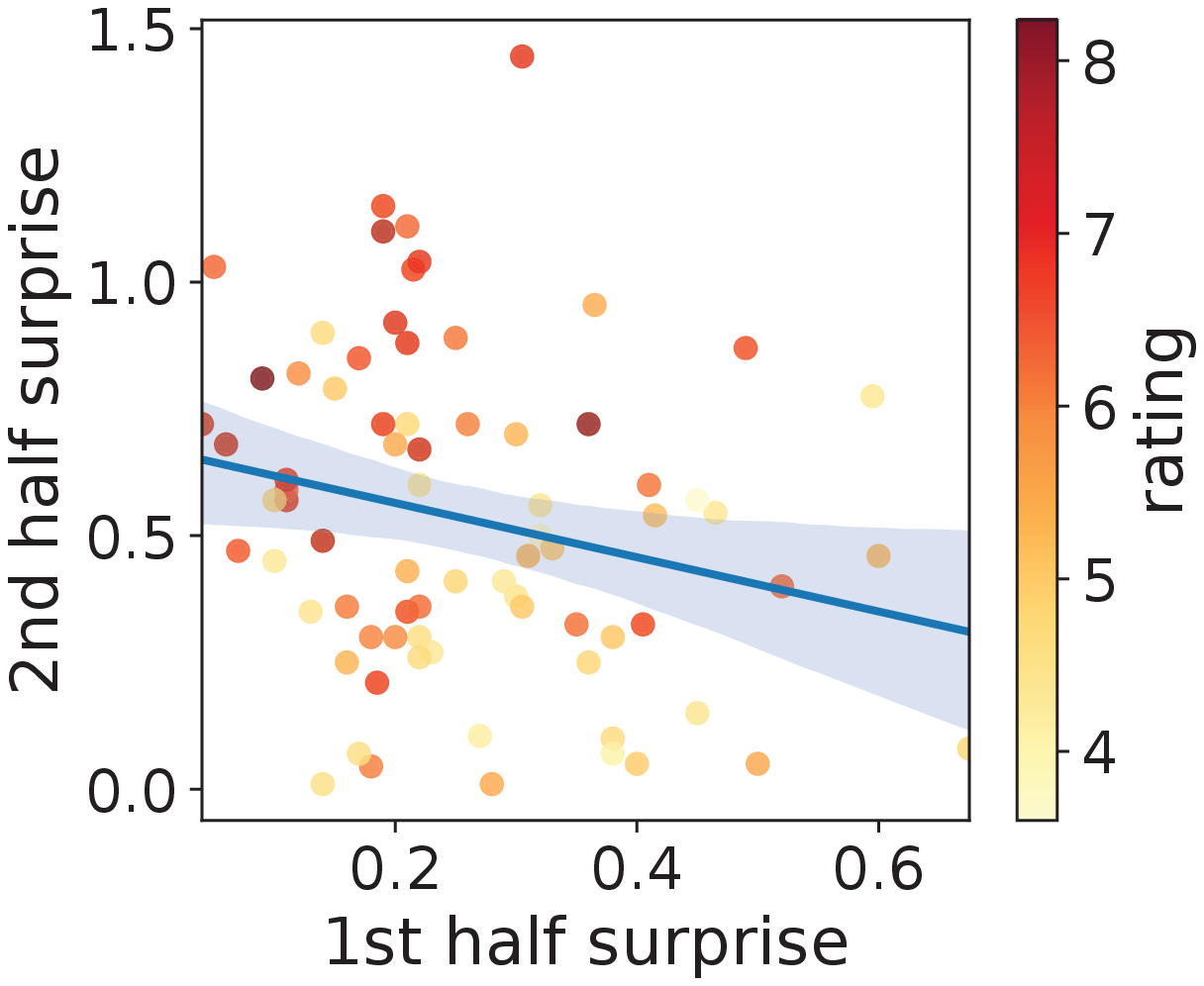}\label{result:1halfvs2half}}
  \subfigure[Peak vs. end]{\includegraphics[width=.48\linewidth]{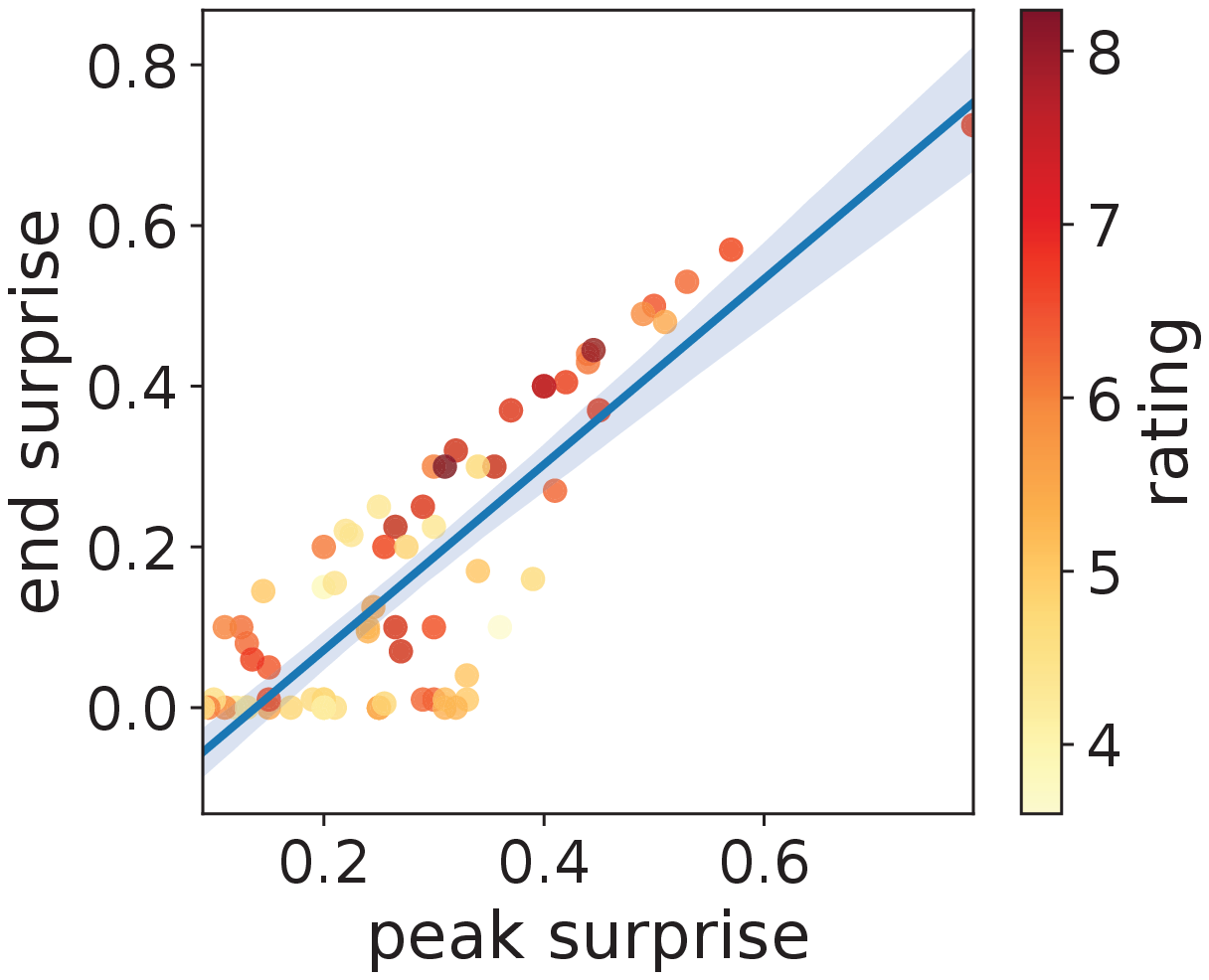}\label{result:peakvsend}}
  \caption{Relationship between surprise in different time: Each point represents a game and is colored by the game's average rating. The figures also show the linear regression lines. }
  \label{result:twopartvs}
\end{figure}

Figure~\ref{result:1halfvs2half} shows a scatter plot of the surprise in the first half and second half of each match.  The points are colored by how exciting the match was, measured by the match's average rating.  We can see that these values are negatively correlated. Figure~\ref{result:peakvsend} is a scatter plot of the surprise in the peak time and end time of each match. These values are positively correlated. We also find that peak is end for 19.7\% games. 

\begin{figure}[!ht]\centering
  \includegraphics[width=.96\linewidth]{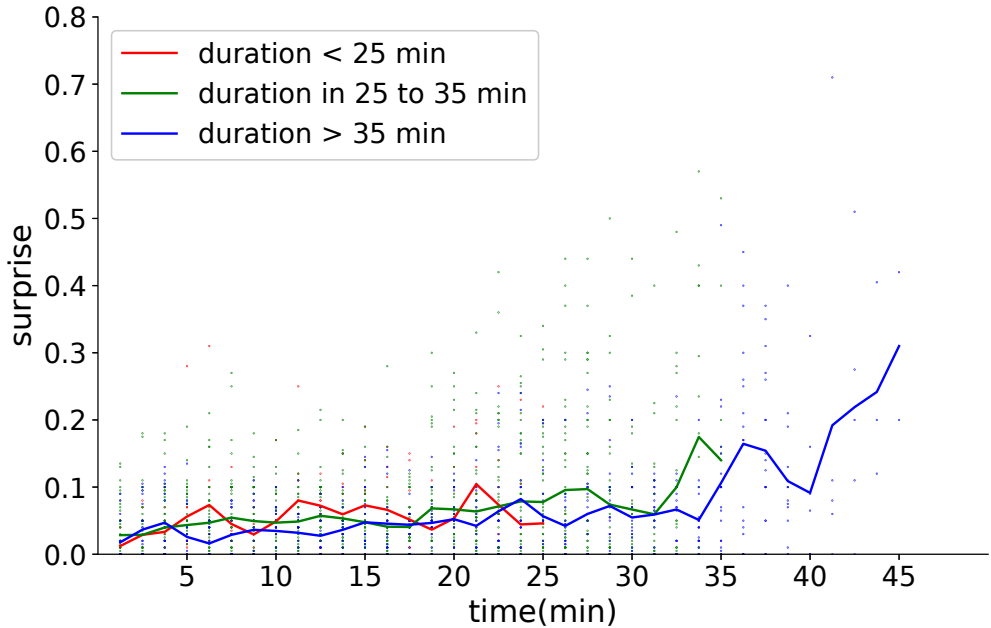}
  \caption{Amount of surprise over time: We discretize time and, at each time, calculate the surprise amount in the time interval of 2.5 minutes centered at that time.  Each dot in the figure represents the surprise amount in a certain game and a certain time interval. The color of a dot shows the duration of the corresponding game. There are three colors in total, red means the corresponding game lasts less than 25 minutes, green means it lasts 25 to 35 minutes, and blue means it lasts more than 35 minutes. The colored lines represent the average surprise amount in games with the same color at a certain time. }
  \label{result:surpDensity}
\end{figure}

Figure~\ref{result:surpDensity} displays the amount of surprise over time.  Short games, tend not to have too much surprise, perhaps, because the teams are unevenly matched.  Long games tend to start off with less surprise, likely as the teams remain evenly matched, but have a substantial amount of surprise toward the end.   

\section{Results}

\begin{figure}[!ht]\centering
  \begin{minipage}{1\linewidth}
  \caption*{\textbf{Uncategorized}}
  \vspace{-.03\linewidth}
  \subfigure[whole game]{\includegraphics[width=.192\linewidth]{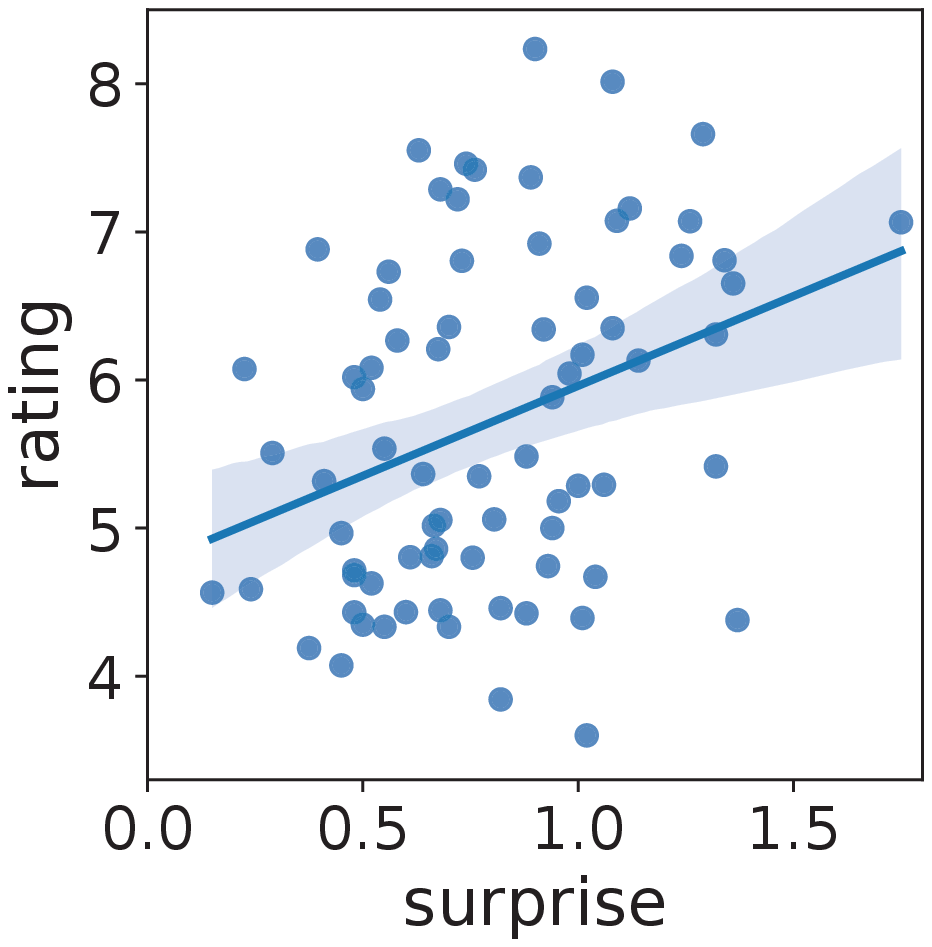}\label{result:fig1:a}}
  \subfigure[1st half]{\includegraphics[width=.192\linewidth]{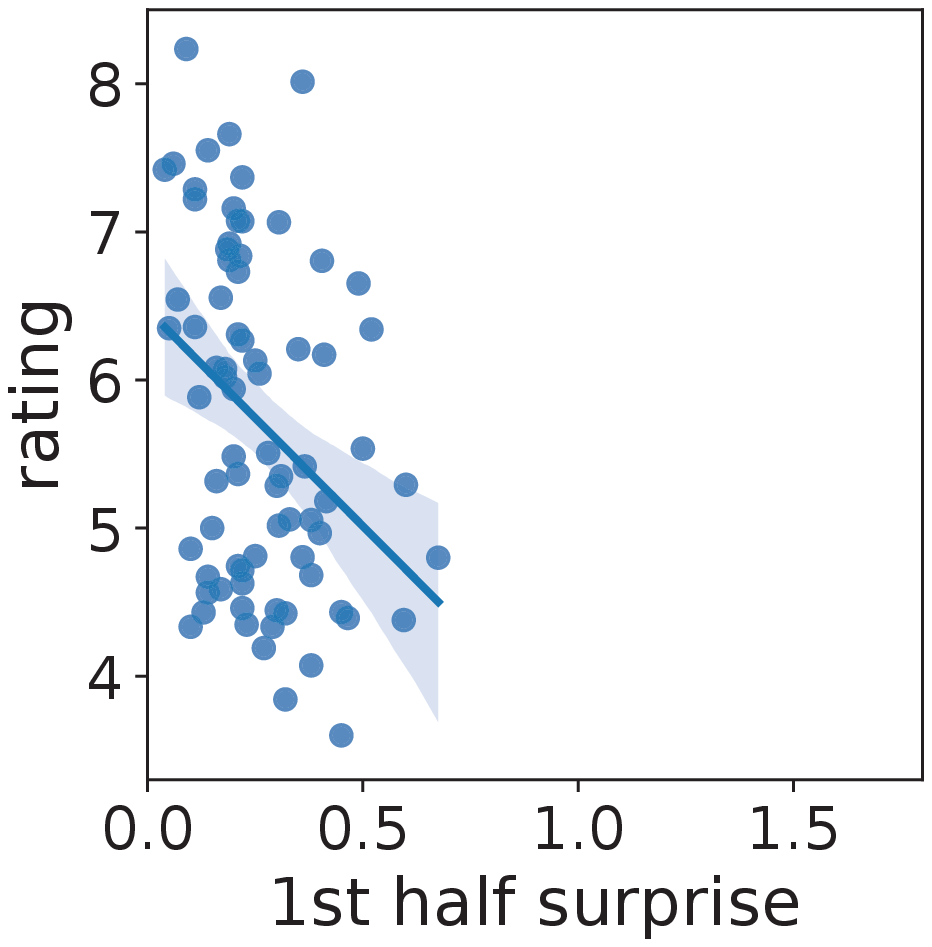}\label{result:fig1:b}}
  \subfigure[2nd half]{\includegraphics[width=.192\linewidth]{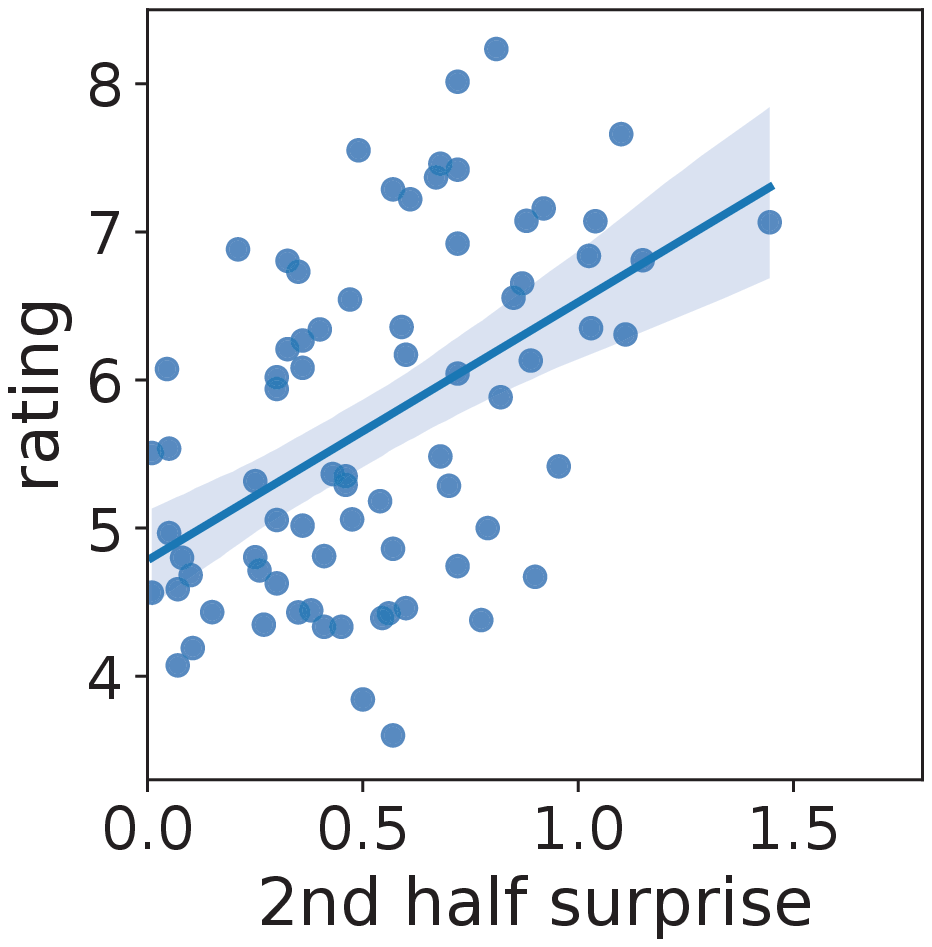}\label{result:fig1:c}}
  \subfigure[peak time]{\includegraphics[width=.192\linewidth]{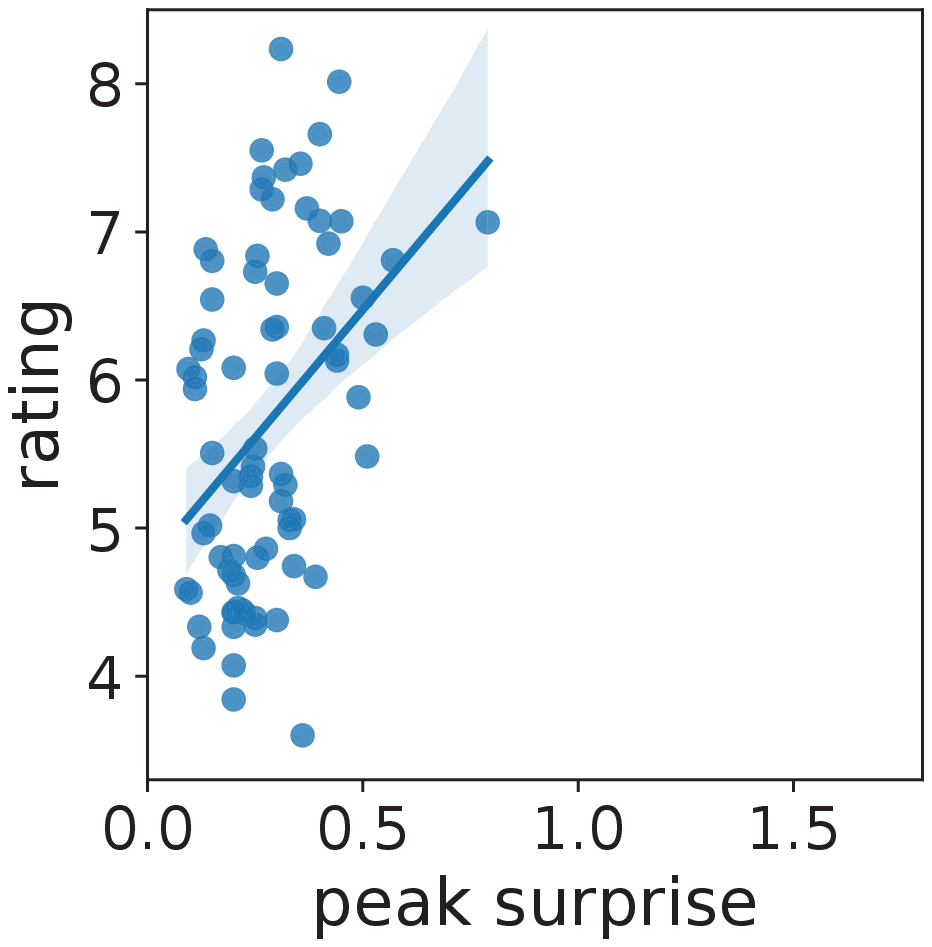}\label{result:fig1:d}}
  \subfigure[end time]{\includegraphics[width=.192\linewidth]{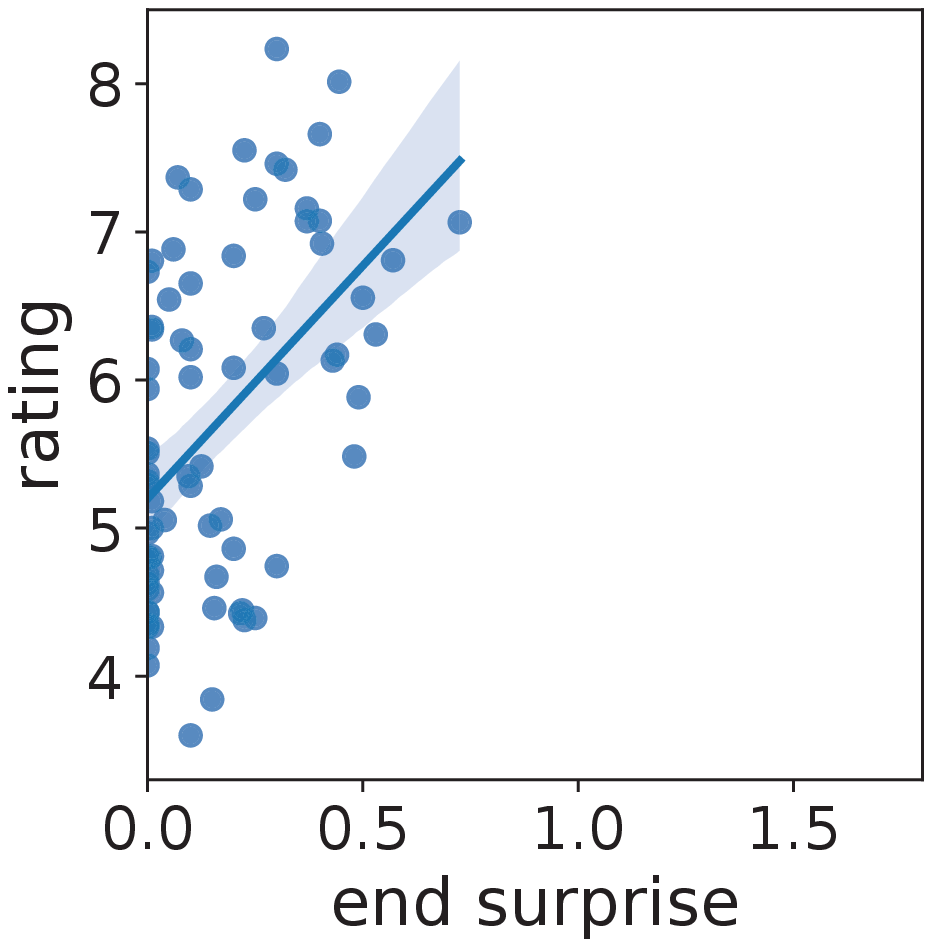}\label{result:fig1:e}}
  \end{minipage}%
  \\
  \vspace{.02\linewidth}
  \begin{minipage}{1\linewidth}
  \caption*{\textbf{Categorized}}
  \vspace{-.03\linewidth}
  \subfigure[whole game]{\includegraphics[width=.192\linewidth]{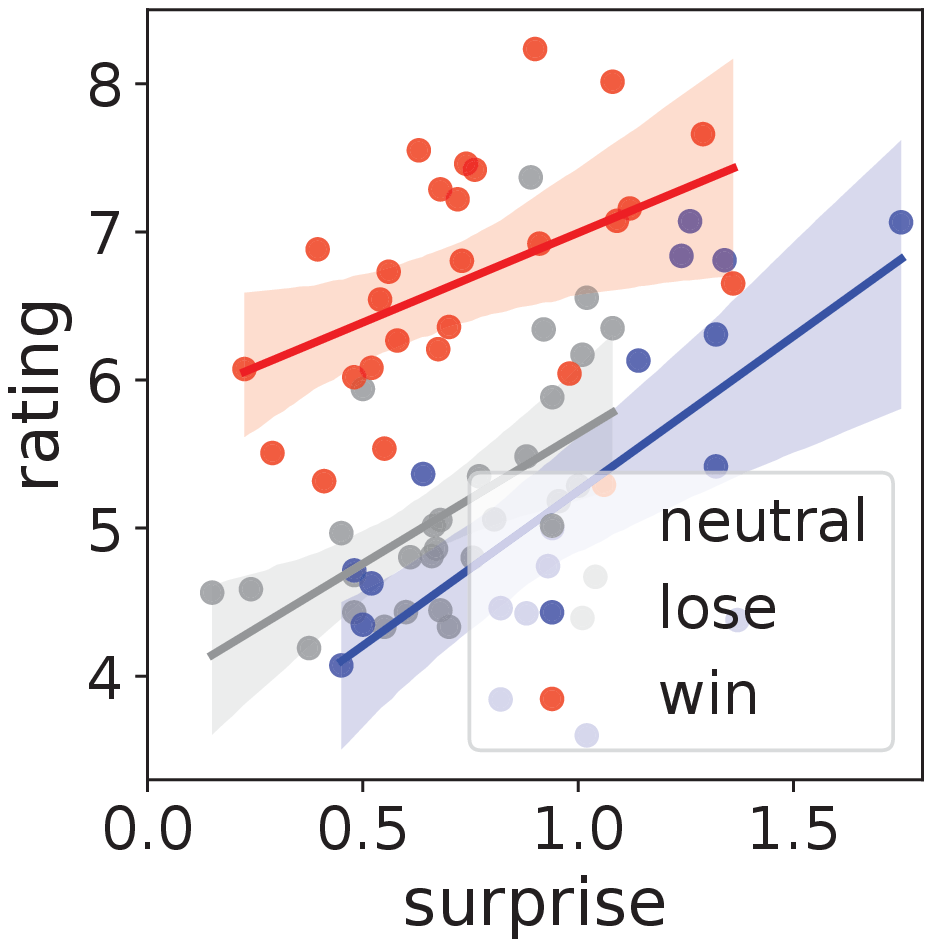}\label{result:fig1:f}}
  \subfigure[1st half]{\includegraphics[width=.192\linewidth]{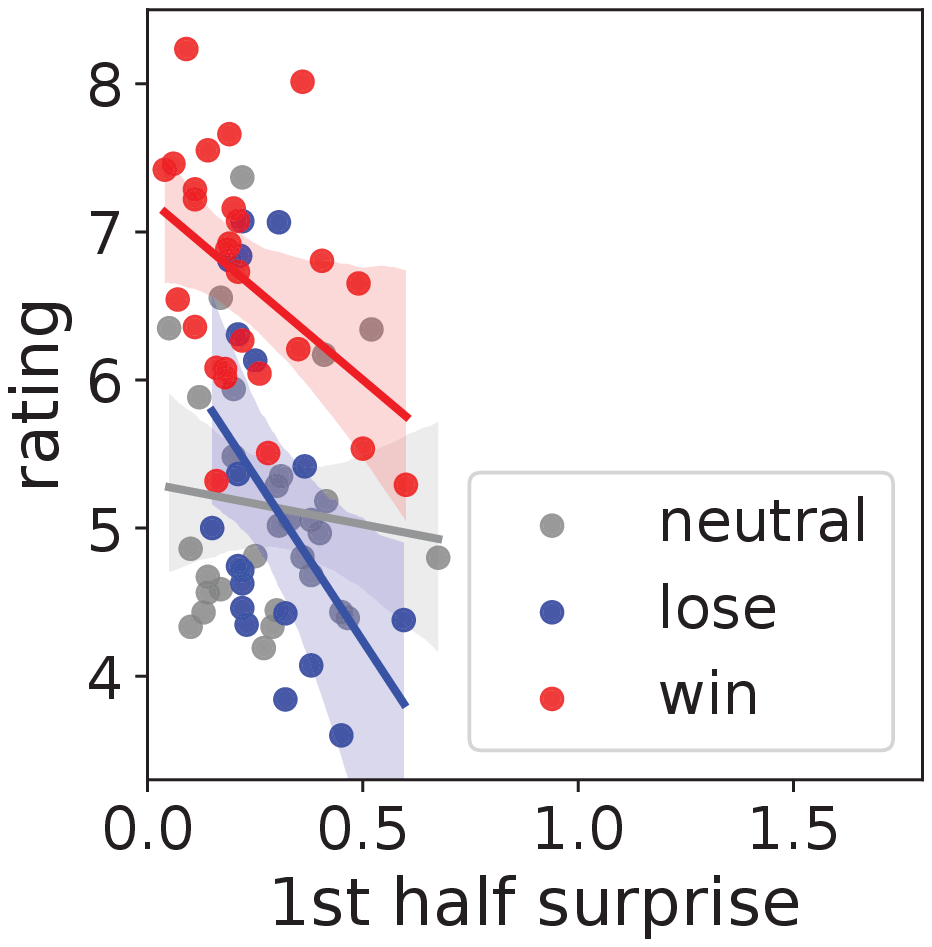}\label{result:fig1:g}}
  \subfigure[2nd half]{\includegraphics[width=.192\linewidth]{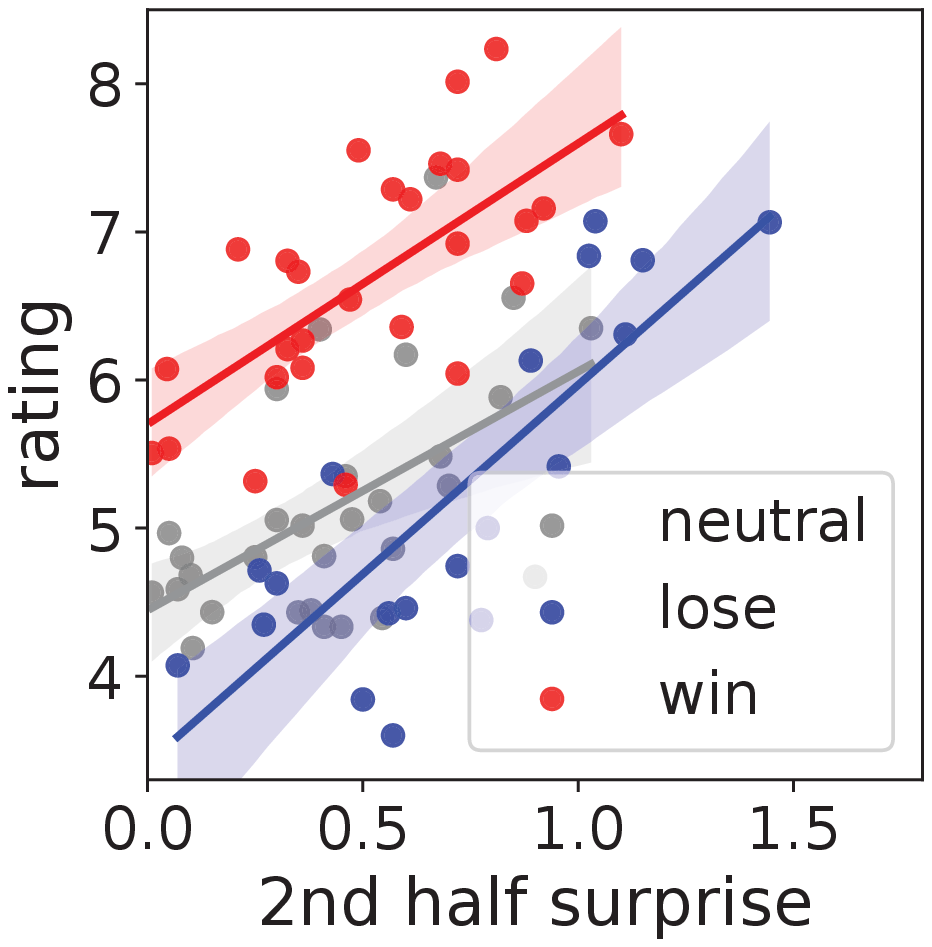}\label{result:fig1:h}}
  \subfigure[peak time]{\includegraphics[width=.192\linewidth]{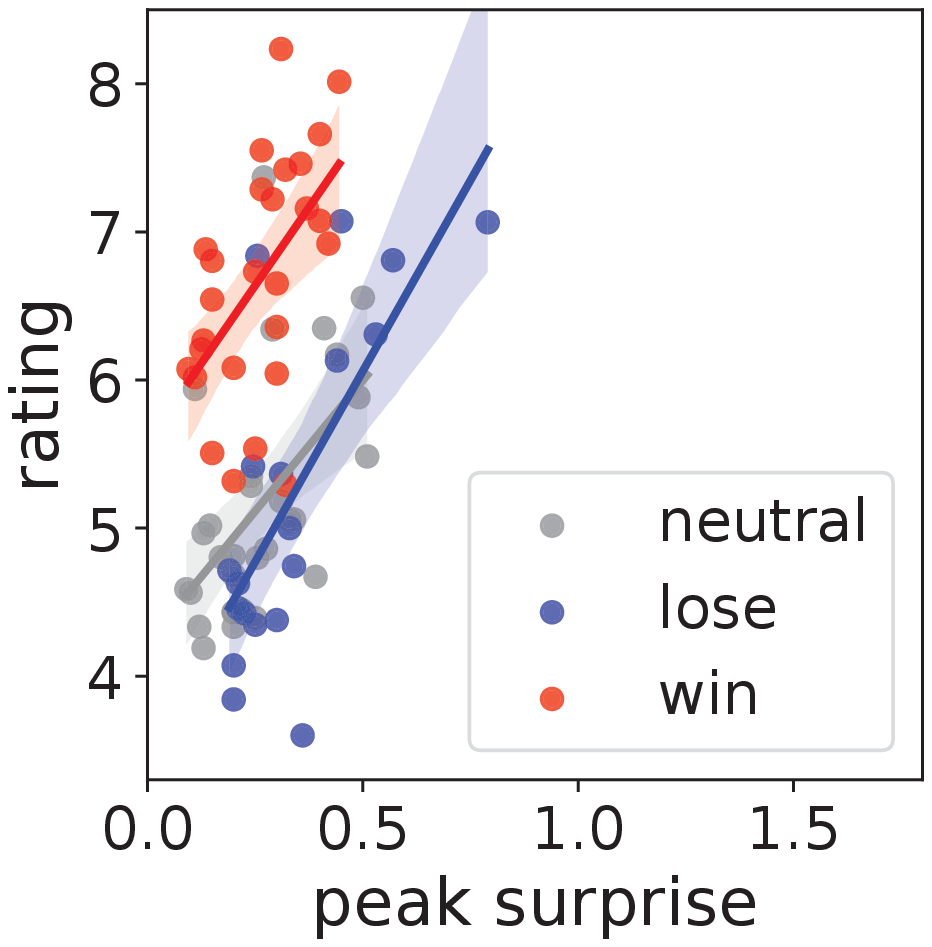}\label{result:fig1:i}}
  \subfigure[end time]{\includegraphics[width=.192\linewidth]{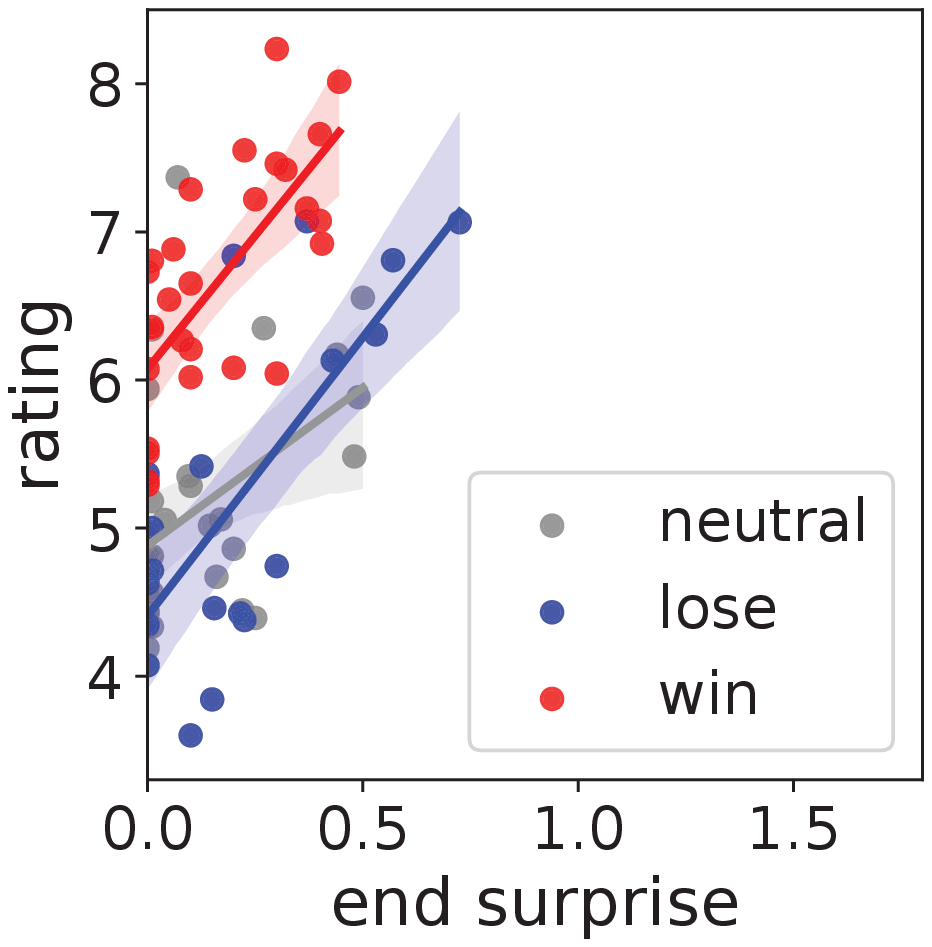}\label{result:fig1:j}}
  \end{minipage}

  \caption{Relationship between surprise amount and rating: In the figures above, each point represents a game and the y-axis is the average rating of the subjects. The x-axis is the surprise amount of the whole game (a,f), the first half (b,g), the second half (c,h), the peak time (d,i), and the end time (e,j) correspondingly. In the second row, the games are classified into three categories. The red points represent the game won by the majority preferred team, the blue dots is the game where the majority preferred team failed, and the gray dots represent the game where no team is preferred by the majority (most of them are neutral.). The second row analyses the relationship between surprise amount and rating using data of only one category. The results are similar to the above row which suggests the robustness of the conclusion. }
  \label{fig:mediancurve}
\end{figure}

First, we analyze the relationship between the subjects' ratings and the amount of surprise in the game. We find that the average rating is significantly positively correlated with the amount of surprise (Figure~\ref{result:fig1:a}, Table~\ref{table:firstandsecond}, Column (1)). We further divide the game into two halves and observe opposite effects between the two time windows. There is a significantly positive correlation between the ratings and the surprise amount in the second half (Figure~\ref{result:fig1:c}, Table~\ref{table:firstandsecond}, Column (3)), while this correlation is negative in the first half (Figure~\ref{result:fig1:b}, Table~\ref{table:firstandsecond}, Column (2)).  This result remains when we regress on both the first half surprise and the second half surprise together (Figure~\ref{result:fig1:b}, Table~\ref{table:firstandsecond}, Column (4)).

\begin{table}[!ht]

\setlength\tabcolsep{3pt}
{
\def\sym#1{\ifmmode^{#1}\else\(^{#1}\)\fi}
\begin{threeparttable}
\begin{tabularx}{1\linewidth}{l*{5}{X}}
\hline\hline
          &\multicolumn{1}{X}{(1)}&\multicolumn{1}{X}{(2)}&\multicolumn{1}{X}{(3)}&\multicolumn{1}{X}{(4)}\\
\hline
Surprise  &   1.214\sym{***}&                  &                  &                  \\
          & (0.399)         &                  &                  &                  \\
1st half &                  &  -2.921\sym{***}&                  &  -2.100\sym{**} \\
surprise          &                  & (0.911)         &                  & (0.846)         \\
2nd half &                  &                  &   1.743\sym{***}&   1.533\sym{***}\\
surprise          &                  &                  & (0.368)         & (0.366)         \\
Constant    &  4.746\sym{***}&  6.473\sym{***}&  4.783\sym{***}&  5.444\sym{***}\\
          & (0.340)         & (0.269)         & (0.227)         & (0.345)         \\
\hline
\(N\)     &       76         &       76         &       76         &       76         \\
adj. \(R^{2}\) &    0.099         &    0.110         &    0.222         &    0.273         \\
\hline\hline
\end{tabularx}
\footnotesize

\end{threeparttable}
}
\caption{OLS regression of surprise and rating in different time periods. The dependent variable is rating score. The independent variable in Columns (1), (2), (3) is surprise, 1st half surprise, 2nd half surprise respectively, and the independent variable in Columns (4) are the 1st half surprise and 2nd half surprise, together. The 1st and 2nd half surprise indicates that the amount of surprise of the 1st and 2nd half of the game. The surprise represents overall amount of surprise in the whole game. Standard errors are reported in parentheses. ***, **, and * indicate statistical significance at the 1\%, 5\%, and 10\% levels, respectively. }
\label{table:firstandsecond}
\end{table}

Importantly, the second half surprise better predicts the average rating than the overall surprise: the coefficient value is 1.743 for second half surprise while it is 1.214 for the overall surprise.  Moreover, the adjusted $R^2$ value is also greater when using second half surprise than when using the overall surprise.  One possibility is that subjects may overweight their watching experiences in the second half of the game. Our result suggests that to optimize the information revelation strategy, the optimization goal should consider time factors and emphasize the later surprise more. 

A possible explanation for this result is the peak-end-effect \cite{kahneman1993more,baddeley1993recency}. That is, people judge an experience mostly based on how they felt at its peak, the most intense point, and at its end, rather than based on the sum of their feeling at all moments of the experience. Thus, we further analyze the effect of the peak surprise and the end surprise in our data (see definition in Section~\ref{sec:ourdata}). Our results show that both of them are highly correlated with the average rating, while the end surprise has the highest correlation (see Table~\ref{table:peakend}). Note that the end surprise explains even more than the second half surprise (the end surprise's adjusted $R^2$ value is $0.232$ which is greater than the second half surprise's adjusted $R^2$ value $0.222$).

\begin{table}[!ht]

\setlength\tabcolsep{3pt}
{
\def\sym#1{\ifmmode^{#1}\else\(^{#1}\)\fi}
\begin{threeparttable}
\begin{tabularx}{1\linewidth}{l*{3}{X}}
\hline\hline
          &\multicolumn{1}{X}{(1)}&\multicolumn{1}{X}{(2)}&\multicolumn{1}{X}{(3)}\\
\hline

Peak surprise &  3.459\sym{***}                &  &        -0.582        \\ 
          & (0.947)                 &         &        (1.637)                \\
End surprise &                  &   3.146\sym{***}               & 3.497\sym{***}   \\
          &                  &    (0.647)              &     (1.183)      \\

Constant    &  4.746\sym{***}&  5.200\sym{***}&  5.304\sym{***} \\
          & (0.290)         & (0.156)         & (0.335)         \\
\hline
\(N\)     &       76         &       76         &       76          \\
adj. \(R^{2}\) &    0.141         &    0.232         &    0.223              \\
\hline\hline
\end{tabularx}
\end{threeparttable}
}
\caption{OLS regression of peak-end surprise and rating. The dependent variable is rating score. The independent variable in Columns (1), (2) is peak surprise, end surprise, respectively. The independent variables in Column (3) are peak surprise, end surprise, together. The peak surprise indicates the  amount of surprise in peak time. The end surprise indicates the amount of surprise generated in the last 2.5 minutes. Standard errors are reported in parentheses. ***, **, and * indicate statistical significance at the 1\%, 5\%, and 10\% levels, respectively. }
\label{table:peakend}
\end{table}

\begin{table}[!ht]

\setlength\tabcolsep{3pt}
{
\def\sym#1{\ifmmode^{#1}\else\(^{#1}\)\fi}
\begin{threeparttable}
\begin{tabularx}{1\linewidth}{l*{4}{X}}
\hline\hline
          &\multicolumn{1}{X}{(1)}&\multicolumn{1}{X}{(2)}&\multicolumn{1}{X}{(3)}&\multicolumn{1}{X}{(4)}\\
           &\multicolumn{1}{X}{all}&\multicolumn{1}{X}{win game }&\multicolumn{1}{X}{lose game}&\multicolumn{1}{X}{neutral game}\\
\hline
Surprise  &   1.692\sym{***}& 1.211\sym{***}&   2.088\sym{***}&   1.760\sym{***}\\
     &   (0.293)   & (0.489)         & (0.557)         & (0.507)  \\     
Win  &  1.498\sym{***}&  &  & \\
     &   (0.198)  &        &   &     \\
Lose   &   -0.376&  &  & \\
      &  (0.232)    &  &      & \\
Constant  &3.928\sym{***}  &   5.783\sym{***}&  3.162\sym{***}&   3.879\sym{***}    \\
         &  (0.251)&  (0.389)         & (0.585)         & (0.385)\\
\hline
$N$    &76 &       27         &       19         &       30         \\
adj. $R^{2}$ &   0.575 &   0.165      &    0.420        &     0.276         \\
\hline\hline
\end{tabularx}
\footnotesize

\end{threeparttable}
}
\caption{OLS regression of surprise and rating in different games. Column (1) is the pooling result. Column (2) refers to games where the majority preferred team wins. Column (3) refers to games where the majority preferred team loses, and Column (4) is for games where no team is preferred by the majority. The dependent variable is the rating score. The independent variable in Column (1) is the amount of surprise, a dummy for winning (losing), i.e.,  whether the game is won (lost) by the majority preferred team. The independent variable in Columns (2), (3), (4) is the amount of surprise, respectively. Standard errors are reported in parentheses. ***, **, and * indicate statistical significance at the $1\%$, $5\%$, and $10\%$ levels, respectively.}
\label{result:table2}
\end{table}

Second, we observe a salient increase (decrease) in ratings for subjects whose preferred team wins (loses). In a game with audiences whose preferences are homogeneous, e.g., a popular team vs. an unpopular team, such individual rating biases lead to an unfairly high (low) average rating for a game depending on the outcome of the game. Therefore, we separate games into three categories: win, lose and neutral. The win (lose) category includes  games where the winning (losing) team was preferred by a majority of subjects.  The neutral category consists of games where neither team was preferred by the majority (recall that subjects can also be neutral in their team preference). The results are in Figure~\ref{result:fig1:f} to \ref{result:fig1:j} and Table~\ref{result:table2}. Again, we observe similar results across all three categories of games. 

In addition to the amount of surprise, we also explore other factors that may affect the audience's average rating. \begin{description}
    \item[\textbf{Comeback size.}] It is defined as one minus the minimum winning probability of the winner during the game. This feature characterizes how big the surprise of the outcome is. The coefficient value is $1.737$ and the adjusted $R^2$ value is $0.029$.
    \item[\textbf{Number of leader changes.}] It is defined as the number of times when the team with more than 50\% winning probability changes. This feature characterizes the team with advantage changes. The coefficient value is $-0.677$ and the adjusted $R^2$ value is $0.017$.
    \item[\textbf{Rubbish time.}] Given a threshold $p$, the rubbish time is defined as the proportion of time period between the last time that the winner's winning probability $\geq p$ and the end of the match. $p$ is a parameter from $0.5$ to $1$. This feature characterizes the unsurprising time before the end. Intuitively, rubbish time is correlated with the end surprise and negatively correlated with the rating. Our results show that $p=0.7$ has the best performance which has the coefficient value as $-1.524$ and the adjusted $R^2$ value as $0.175$.
\end{description}  Among the above three factors, the rubbish time is the most relevant factor but is still less effective than the end surprise.

\section{Related Work}

\paragraph{Surprise vs. perceived quality.} Starting from \citet{ely2015suspense}, a growing literature examines the relationship between the surprise and the perceived quality in different games, such as tennis games  \cite{bizzozero2016importance}, soccer games \cite{buraimo2020unscripted} and rugby games \cite{scarf2019outcome}. 

However, instead of eliciting belief curves, this literature tyfiguresally constructs them from existing data.
For example, \citet{bizzozero2016importance} model the probability of a certain side winning by explicitly using tennis's scoring systems.  Similarly, \citet{buraimo2020unscripted} use an in-play model which additionally exploits the information on team strength which is embedded in the pre-match odds. \citet{buraimo2020unscripted} and \citet{scarf2019outcome} both use the Poisson model to estimate the number of goals scored by the participating teams in order to calculate the probability of the final outcome of the game. \citet{LUCAS201758} analyze the tweets during the World Cup and use the emotional changes to measure the surprise. In contrast to these studies which estimate  surprise using statistical models, our study measures the perceived surprise amount by dynamically eliciting subjects' beliefs. 

These works also use different proxies for perceived quality. \citet{buraimo2020unscripted} analyze the relationship between surprise and the real-time audience size for both halves of soccer games.  Instead, we focus on the relationship between surprise and the overall rating. 
Thus, prior studies do not observe how the timing effects of surprise affect perceived quality. 

In addition to surprise, massive literature also studies suspense, which is defined as how much the belief curve is expected to move in the very near future.  Since measuring suspense requires the ability to predict what might happen in the near future, the analysis for suspense is beyond the scope of the paper. In the future, it might be possible to learn a model from the data which enables the analysis of suspense. 

\paragraph{Prediction markets and polls.} Prediction markets are designed to elicit continuously updated forecasts about uncertain events. Prior studies have proved that prediction market can outperform internal sales projections~\cite{plott2002information}, journalists’ forecasts of Oscar winners~\cite{pennock2001extracting}, and expert economic forecasters~\cite{gurkaynak2006macroeconomic}. 

In prediction polls, forecasters express their beliefs by answering questions like "how likely is this event?". 
In both prediction markets and polls, forecasters can update their predictions whenever they wish. In contrast to prediction markets, in prediction polls, forecasters update their predictions individually. \citet{rothschild2011forecasting}'s work shows that using prediction polls in elections can achieve better accuracy than opinion polls.

A few studies have compared the performance of prediction markets and polls, though there is no conclusive answer \cite{goel2010prediction,rieg2010forecasting,atanasov2017distilling}. 
Both \citet{goel2010prediction}  and \citet{rieg2010forecasting} find no significant differences between these two methods. \citet{atanasov2017distilling} find that the aggregation rules in prediction polls affect its accuracy level. For example, simply averaging all polls performs worse than a prediction market, while weighting the polls properly leads to a better performance than a prediction market. Our elicitation method is more similar to prediction polls. 

\section{Conclusion}

We study the relationship between a game's surprise amount and its perceived quality.  We develop a platform that collects audiences' real-time beliefs and ratings of the LOL S10. Our empirical analysis suggests that the level of surprise in the later time of a game has a stronger impact on subjects' ratings. This indicates that the audience would prefer surprise to occur in the end of a game.  A future direction is to define a new perceived quality model that considers time factors and theoretically analyze the optimal way to reveal information over time in this model.  Future work could similarly optimize suspense.

Moreover, we expect that belief polls could be embedded in live streaming games as an entertainment feature. Last but not least, we can collect other information such as the text in bullet comments to better construct the information flow.

\section*{Acknowledgments}
We would like to thank our anonymous reviewers for their insightful suggestions which substantially improved the paper. We also thank Kening Ren, Jialun Yang, Xinlun Zhang, Fan Yan and Zheng Zhong for their help and useful discussions. Finally, we thank our participants in our LOL S10 study for their time, attention and effort. Some of the figures are generated with Python Matplotlib~\cite{Hunter:2007}. 


\newpage
\bibliographystyle{plainnat}
\bibliography{ref}

\end{document}